\documentclass{article}

\usepackage{PRIMEarxiv}

\usepackage[utf8]{inputenc} % allow utf-8 input
\usepackage[T1]{fontenc}    % use 8-bit T1 fonts
\usepackage{hyperref}       % hyperlinks
\usepackage{url}            % simple URL typesetting
\usepackage{booktabs}       % professional-quality tables
\usepackage{amsfonts}       % blackboard math symbols
\usepackage{nicefrac}       % compact symbols for 1/2, etc.
\usepackage{microtype}      % microtypography
\usepackage{lipsum}
\usepackage{fancyhdr}       % header
\usepackage{graphicx}       % graphics
\graphicspath{{media/}}     % organize your images and other figures under media/ folder

\usepackage[numbers,square,sort&compress]{natbib}
\usepackage{url,hyperref,lineno,microtype,subcaption}
\usepackage[onehalfspacing]{setspace}
\usepackage{lscape}
\usepackage{amsmath}

\fancypagestyle{specialfooter}{%
  \fancyhf{}
  
  \fancyfoot[R]{\textit{Preprint submitted to Frontiers in Physics \hfill \today}}
}

\title{Wealth distribution in villages. Transition from socialism to capitalism in view of exhaustive wealth data and a master equation approach.
}

\author{
  Istv\'an Gere\,$^{1}$, Szabolcs Kelemen\,$^{1}$,  Tam\'as S. Bir\'o\,$^{1,2}$ and Zolt\'an N\'eda\,$^{1,*}$ \\
  $^{1}$Babe\c{s}-Bolyai University, Dept. of Physics, Cluj-Napoca, Romania \\
  $^{2}$Wigner Research Center for Physics, Budapest, Hungary \\
}

\begin{document}
\maketitle
\thispagestyle{specialfooter}

\begin{abstract}
Socio-economic inequalities derived from an exhaustive wealth distribution is studied in a closed geographical region from Transylvania (Romania). Exhaustive wealth data is computed from the agricultural records of the Sancraiu commune for three different economic situations.  The gathered data is spanning two different periods from the communist economy  and the present situation after 31 years of free market economy in Romania. The local growth and reset model based on an analytically solvable master equation is used to describe the observed data. The model with realistically chosen growth and reset rates is successful in describing both the experimentally observed distributions and the inequality indexes (Lorenz curve, Gini coefficient and Pareto point) derived from this data. The observed changes in these inequality measures are discussed in the context of the relevant socio-economic conditions.
\end{abstract}

% keywords can be removed

\section{Introduction}

Fascinating statistics related to cities or smaller settlements have been intensely studied by physicists in the last few decades. On the data mining and analyzing side, the studies coming from physics revealed many  universalities and scaling laws. Modeling was done with simple physics inspired models that were able to prove by their success the relevancy of some socio-economic processes in the investigated phenomenon. Numerous universalities have been revealed and successfully modeled \cite{scalingincities,scalingincities3,scalingincities4,scalingincities5,scalingincities6, scalingincities7,scalingincities8}.

It has been shown that despite the large differences in their population,  cities are all qualitatively similar from the point of view of most of the sociometric indices. Many of the urban sociometric measures, such as length of roads, total income, GDP, total wages, gross urban product etc... are scaling with the population of the settlement \cite{scalingincities, scalingincities4, scalingincities6}. Based on the value of the scaling exponent, for cities from the USA, Germany and China many of these urban indicators were categorized by Bettencourt et al. in three regimes: sublinear, linear and superlinear \cite{scalingincities4}. An elegant explanation for the reason why larger cities are growing on the expense of smaller ones were offered by interpreting this different scaling regimes.

The social inequality in a settlement's population is also affected by the size of the population, however the methodology of measuring and modeling social inequalities is more complex than the one used for most of the other urban indices mentioned above. Social inequalities have been studied beginning from the early days of Economics and presently is one of the main direction in Econophysics \cite{chakraborti}. Everybody is aware of Vilfredo Pareto's seminal works  \cite{pareto}, revealing the universality of the fat-tailed distributions in income and wealth and the related 80-20$\%$ law. 
Econophysics targets the problem of inequalities in a society starting from the experimental distribution of income or wealth of individuals (or groups of individuals) of a well-delimited population \cite{yakovenko}. 

The most commonly accepted measure that characterizes the inequality in a society is the Gini index \cite{yakovenkko09}. It has been shown that the value of the Gini index grows with the size of the city, meaning that inequality is more pronounced in larger cities, than in case of smaller ones \cite{scalingincities7}. Recently, a more profound study focusing on the effect of city size on the income distribution was published by E. Heinrich Mora et. al.  \cite{scalingincities3}, showing that the income does not scale in the same manner for all regions of the society. The scaling is sub-linear for the lower regions of the society and scales super-linear in case of the higher regions. (\textit{This result is in agreement with the wealth data from a small settlement (about 1000 households) processed in this article.}) The results of Mora et. al. illustrates nicely the effect of increasing inequality with the city size.
\vspace{5pt}
% Not only the way how social inequality scales with the size of the population is fascinating, it is also fascinating to understand the underlying mechanism leading to the unusual shape of the distribution of such socioeconomic measures like the income or wealth./

\noindent
Not only the fat-tail of social inequalities, but also the understanding of the underlying mechanism leading to the unusual distribution of such socioeconomic measures like income or wealth are fascinating questions. It is known nowadays that the distribution of income and wealth is not a simple one, different type of distributions apply to different regions of income or wealth. The richer end of both distributions can be described with the power-law type distribution \cite{pareto,yakovenko}. The low and middle classes of income and wealth can be described better with an exponential distribution. In case of wealth a third region should be considered as well, the region of negative wealth (debts), which is also characterized by an exponential trend, different from the one which applies for the low and middle classes \cite{ZNeda2}.

 Modeling the experimental data is done by many different methods, ranging from simple mean-field type models, to more complex models involving network-science approach or agent based computational simulations \cite{yakovenko, Cui, Cardoso, Jongsoon, Lim, levy,jones,sorger, Clementi, Bouchaud, Chatterjee, Oliveira, ZNeda_gambler, ZNeda_family, ZNeda1, Ciesla}. 

Recently, we proposed a simple master-equation approach based on a Local Growth and Global Reset (LGGR) \cite{BNT}process to describe in a similar manner the distribution of both income and wealth \cite{ZNeda1, ZNeda2}. The novelty of this approach is also its ability to describe all income and  wealth regions in a unified manner,  proposing a compact form of the distribution function for all income/wealth categories, capturing the regions of the negative wealth as well.

Although income and wealth seems to be related and can be modeled with similar approaches, from data mining side there are big differences between them. While incomes are present in many electronically available data (taxation for example), wealth is  more difficult to quantify and measure directly and raises many private issue problems. Wealth is usually estimated via some proxies and it is  hard to find exhaustive data for a well-delimited  and statistically significant community. 
In our previous studies targeting income distributions \cite{income_n,ZNeda1} we have shown the advantages of having a complete statistics of income (exhausting data) in a population for many consecutive years. Such data allows not only to 
test the statistics offered by the model, but it also allows verification of some hypothesis used in modeling the dynamics of the system. 

Exhaustive real-world wealth data are rare to find in the literature and as a consequence, these could be extremely precious from the point of view of modeling studies on socio-economics problems targeting social-inequalities. Here we construct such an exhaustive wealth database extracted from taxation and ownership data that is available at the mayor's office in a Romanian village community, \textit{comuna Sâncraiu,} (\textit{Kalotaszentkir\'aly}), having about 1000 households. The database includes data from three different years: 1961, 1989 and 2021. The real beauty of this data is that the studied years are relevant snapshots for three very different political periods,  influencing in a different manner private wealth.  The year 1961 is exactly the year before collectivization, when the lands given to the peasants by the communists after 1947 were again confiscated and collectivized. 1989 was the last year of the communism regime in Romania and finally 2021 is the present situation after more than 30 years after the fall of communism offering the picture of the effects of the free market economy in Romanian villages.  

Besides presenting and discussing the relevant wealth distributions derived from the data, our additional aim is also to validate once again the modeling power of the LGGR model \cite{ZNeda2} in describing social inequalities. Possessing exhaustive wealth data for three different economic periods in Romania we are also able to show how social inequalities varied in these turbulent times and how the growth and reset rates should be adjust in the LGGR model to account for the measured wealth-distributions. Besides the relevant distribution functions we discuss and compare the Lorenz curves, Pareto points and the value of the Gini indices for all the studied periods. 

The rest of this manuscript is organized as follows: (1) we present and discuss the data; (2) the LGGR modeling framework  is briefly discussed; (3) the LGGR model is applied to the obtained wealth data by using different growth and reset rates; (4) the obtained results are discussed in view of socio-economic inequalities, and (5) finally we summarize our findings.

\section{The exhaustive wealth data}

Wealth is hardly quantifiable since it is composed by all valuables that a person possesses. Previous studies regarding wealth distributions were based either on the inheritance tax data \cite{yakovenko_w}, estimated wealth data provided by media companies (Forbes) or organizations such as the one managing the World Inequality Database (WID) \cite{ZNeda1}. A first shortcoming of these data is their incomplete nature and a biased sampling: targeting the aging part of the population or the rich, for example. Analysis performed on exhaustive wealth data is lacking from the literature to our knowledge. Importance of such exhaustive data was recognized in our previous works on income distribution. The exhaustive data for income derived from a anonymized social-security database in ten consecutive years  in the Cluj county (Romania) \cite{income_n,ZNeda2}, allowed to model realistically the dynamics of the income and to describe successfully its distribution.

Here we aim to understand wealth distribution by first constructing an exhaustive data collection for a delimited geographical territory in Romania. The used data and the compiling method that is presented in the following allowed to infer the wealth of each household in a commune of Cluj county. Wealth datasets from three radically different economic life situations of the villages from Romania are considered: 
\begin{itemize}
 \item \textbf{1961} The year under the communist regime where most of the agriculture destined land was confiscated and collectivized, following the  land reform from 1945.
 \item \textbf{1989} The last year of the communist regime, when most of the private property is already abolished, leaving households with limited valuables.
 \item \textbf{2021} The current year, after 31 years the abolishment of the communism and transition to a free market economy.  
\end{itemize}

\begin{figure}[!ht] 
\begin{center}
\includegraphics[width=13cm]{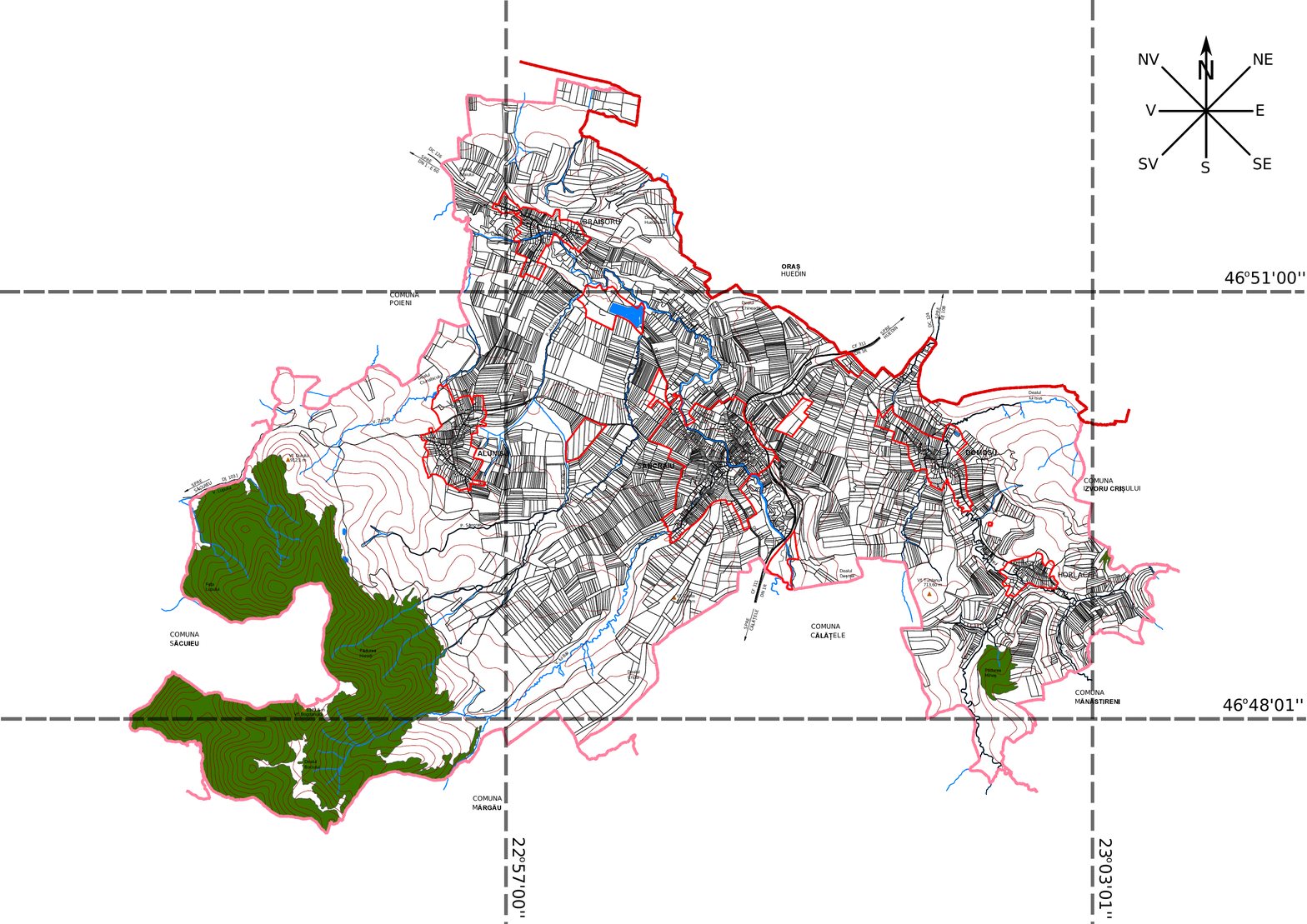}
\end{center}
\caption{The recent publicly available map of commune Sancraiu. The map presents the boundaries of the commune, the neighboring communes and city, along with the constituent villages, intra-village territory, and agricultural  land parcels. Courtesy of Sancraiu Mayor's office.}
\label{fig:Sancariu_osm}
\end{figure}

% (Asterisk: A commune is an administrative subdivision in the state of Romania.) 
Following an agreement between the Babes-Bolyai University of Cluj, and commune Sancraiu (in Hungarian: Kalotaszentkiraly$-$Zentelke) records regarding the wealth of households in the commune were obtained in anonymized format. The commune is a well-delimited region in Transylvania, consisting of five villages: Alunisu (Magyarokereke), Braisoru (Malomszeg), Domosu (Kalotadamos), Horlacea (Jakotelke) and the eponymous Sancraiu. The region and these villages can be located on the map of Figure \ref{fig:Sancariu_osm}. The current population of the commune is $1628$ inhabitants according to the census from $2011$. The $1956$ census enumerates $3557$ inhabitants \cite{census_older} and  the census from $1992$ records a population of $2.053$ individuals \cite{census_older}.

The wealth component data for $1961$ and $1989$ was obtained by digitalizing in an anonymous manner the agricultural registers of the mayor office (Agricultural Register of Sancraiu commune for 1959-1960-1961,  Agricultural Register of Sancraiu commune for 1959-1960-1961), while for the year $2021$ we used the anonymized taxation database for land and buildings.

The "agricultural register" is a complete agricultural record kept by the local authorities. These records contain each household in the commune along with the owned land area, house area, auxiliary building area (such as barns etc.), along with the number of owned livestock in different category (horses, cows, sheep, etc.). As the commune is located historically in an agricultural area, we assumed  that before $1990$ the agricultural records contain the bulk of the valuables that a household owns. In the data there is no estimation of the total wealth of the households, this can be approximated only by transforming all the mentioned valuables with appropriate weights in a hypothetical good. With a realistic weighted sum of all recorded valuables one can create thus a reasonable proxy, that estimates the wealth of an individual household. The obtained wealth is thus a quantity  expressed in arbitrary units ($a.u.$), and their values cannot be compared between the different years, because the wealth components and also their share in the total wealth is changing in time. In the following we describe each dataset in detail, together with the estimation method for the total wealth.

\subsection{Wealth data for 1961}
The Agricultural Registers 1959-1960-1961 contain information about each household in the commune. The total number of individual households that are recorded are $1133$. For each household the following data are used to construct a proxy for the wealth:
\begin{itemize}
 \item $A_L$ -  total land owned in hectares ($ha$). All types of land are included. The weight parameter  will be denoted by $P_{A_L}$. 
 \item $A_{Bh}$ - area of the house owned in $m^2$. Weight parameter: $P_{A_{Bh}}$. 
 \item $A_{Ba}$ - total area of the auxiliary buildings (such as barns, etc.). Weight parameter: $P_{A_{Ba}}$.
 \item $N_C$ - number of cows owned. Weight parameter: $P_{N_{C}}$.
 \item $N_{WB}$ - number of domestic water buffaloes owned. Weight parameter: $P_{N_{WB}}$.
 \item $N_H$ - number of horses. Weight parameter: $P_{N_H}$. 
 \item $N_P$ - number of pigs. Weight parameter: $P_{N_P}$.
 \item $N_S$ - number of sheep. Weight parameter: $P_{N_S}$.
 \item $N_G$ - number of goats. Weight parameter: $P_{N_G}$.
\end{itemize}
The dataset is diverse enough to allow a robust approximation of the wealth. The total wealth of household $i$ is calculated as a simple weighted sum of these valuable categories
% \begin{equation}
\begin{multline}
 W_i = A_{Li} P_{A_L} + A_{Bhi} P_{A_{Bh}} + A_{Bai} P_{A_{Ba}} + N_{Ci} P_{N_C} + N_{WBi} P_{N_{WB}} + \\ + N_{Hi} P_{N_H} + N_{Pi} P_{N_P} + N_{Si} P_{N_S} + N_{Gi} P_{N_G},
\end{multline}
with the $P_i$ weight factors satisfying the normalization: $\sum_i P_i=1$.

% \end{equation}
Since the values of the weighting parameters are largely disputable, we estimate the individual wealth values with 10 different (but realistic) parameter sets, allowing also an estimation for the uncertainty of the wealth estimation method. The chosen parameter sets are presented in Table \ref{tbl:Params61}. To fully understand the effect of  different parametrizations,  we calculate also the share of a single "category" from the total wealth in the commune. 
The total wealth is calculated as $W_{tot} = \sum_i W_i$. The share of the category $C$ with weighting parameter $P_C$ in th etotal wealth (this can be the land $A_L$, house $A_{Bh}$ etc.) in percents will be: 
\begin{equation}
 S(C) = \frac{P_C\sum_i C_i}{W_{tot}} \times 100 \%
 \label{shares}
\end{equation}
We present the shares by categories for the parameter sets from Table \ref{tbl:Params61}, in the Table \ref{tbl:Shares 61}. The data from this table indicates that the used parameter sets cover a wide range of acceptable methods to compound a wealth proxy and the uncertainty estimated from here is realistic. 

The experimental wealth distribution $\rho(w)$ is calculated for each parameter set. The probability density function is built up from the individual household wealth values with the histogram method. For all parameter sets we used the same number of bins for the histogram, however 
the middle point of each bin becomes slightly different, leading to an average value and also a corresponding error bar. The error bars for the 
probability density function is estimated from the different bin intervals and histogram values similarly.  The error bars on the points in the direction of both axes signify the greatest deviation from the mean of the bin. The use of these error bars allows us to represent in a compact manner the ensemble of distributions obtained with the used parameter sets. 
\begin{table}[!ht]
{\small
\begin{center}
\begin{tabular}{|l|l|l|l|l|l|l|l|l|l|}
\hline
P. & $P_{A_L}$    & $P_{A_{Bh}}$    & $P_{A_{Ba}}$   & $P_{N_{C}}$ & $P_{N_{WB}}$ & $P_{N_H}$ & $P_{N_P}$   & $P_{N_S}$    & $P_{N_G}$     \\ \hline
1       & 0.1893 & 0.0068  & 0.0030 & 0.1893  & 0.2158   & 0.3786 & 0.01515 & 0.0015 & 0.0004 \\ \hline
2       & 0.2435 & 0.0062 & 0.0026 & 0.1826  & 0.1753    & 0.3652  & 0.0219 & 0.0022 & 0.0005 \\ \hline
3       & 0.2799 & 0.0058 & 0.0022 & 0.1781  & 0.1481   & 0.3562 & 0.0265 & 0.0026 & 0.0006 \\ \hline
4       & 0.3060 & 0.0055 & 0.0020 & 0.1749  & 0.1285    & 0.3497 & 0.0297 & 0.003 & 0.0006 \\ \hline
5       & 0.3257 & 0.0053 & 0.0018 & 0.1724  & 0.1138   & 0.3449 & 0.0322 & 0.0032 & 0.0007 \\ \hline
6       & 0.3534 & 0.0050 & 0.0016 & 0.1690  & 0.0931   & 0.3380 & 0.0356 & 0.0036 & 0.0007 \\ \hline
7       & 0.3635 & 0.0049 & 0.0015 & 0.1678  & 0.0856  & 0.3355 & 0.0369 & 0.0037 & 0.0007 \\ \hline
8       & 0.3719  & 0.0048 & 0.0014  & 0.1667  & 0.0793  & 0.3334 & 0.0380 & 0.0038 & 0.0007 \\ \hline
9       & 0.3790 & 0.0047 & 0.0014 & 0.1658   & 0.0739  & 0.3317 & 0.0389 & 0.0039 & 0.0008 \\ \hline
10      & 0.3852 & 0.0046 & 0.0014 & 0.1651  & 0.0693  & 0.3301 & 0.0396  & 0.0040  & 0.0008 \\ \hline
\end{tabular}
\end{center}
\caption{Weight parameter sets considered for the data from year 1961.}
\label{tbl:Params61}
}
\end{table}
\begin{table}[!ht]
{\small
\begin{center}
\begin{tabular}{|l|l|l|l|l|l|l|l|l|l|}
\hline
P. & $S(A_L)$   & $S(A_{Bh})$    & $S(A_{Ba})$   & $S(N_{C})$ & $S(N_{WB})$ & $S(N_H)$ & $S(N_P)$   & $S(N_S)$    & $S(N_G)$  \\ 
   & \% & \% & \%  & \% & \% & \% & \% & \% & \% \\ \hline
 1  & 35.3678 & 19.5784 & 10.752  & 5.4663 & 22.9659 & 4.3239 & 1.3365 & 0.203519 & 0.0057 \\ \hline
2  & 44.2776 & 17.3616 & 8.8335 & 5.1325 & 18.1588 & 4.0599 & 1.8823 & 0.2866 & 0.0071 \\ \hline
3  & 50.0034 & 15.937  & 7.6007 & 4.9180 & 15.0695 & 3.8902 & 2.2331 & 0.3401 & 0.0081 \\ \hline
4  & 53.9932 & 14.9443 & 6.7416 & 4.7686 & 12.9169 & 3.7720 & 2.4775 & 0.3773 & 0.0087 \\ \hline
5  & 56.9326 & 14.213  & 6.1086 & 4.6584 & 11.3310  & 3.6849 & 2.6576 & 0.4047 & 0.0092 \\ \hline
6  & 60.9736 & 13.2076 & 5.2385 & 4.5071 & 9.1508 & 3.5651 & 2.9051 & 0.4424 & 0.0098 \\ \hline

7  & 62.4220  & 12.8472 & 4.9266 & 4.4528 & 8.3693 & 3.5222 & 2.9939 & 0.4559 & 0.0101  \\ \hline
8  & 63.6207 & 12.5490  & 4.6685 & 4.4079 & 7.7225 & 3.4867 & 3.0673 & 0.4671 & 0.0103  \\ \hline
9  & 64.6291 & 12.2981 & 4.4514 & 4.3701 & 7.1785 & 3.4568 & 3.1291 & 0.4765 & 0.0104  \\ \hline
10 & 65.4891 & 12.0841 & 4.2662 & 4.3379 & 6.7145 & 3.4313 & 3.1818 & 0.4845 & 0.0106  \\ \hline
\end{tabular}
\end{center}
\caption{Share in the total wealth for different valuable categories. The rows are the results for the considered parameter sets in year 1961}
\label{tbl:Shares 61}
}
\end{table}

\begin{figure}[!ht] 
\begin{center}
\includegraphics[width=10cm]{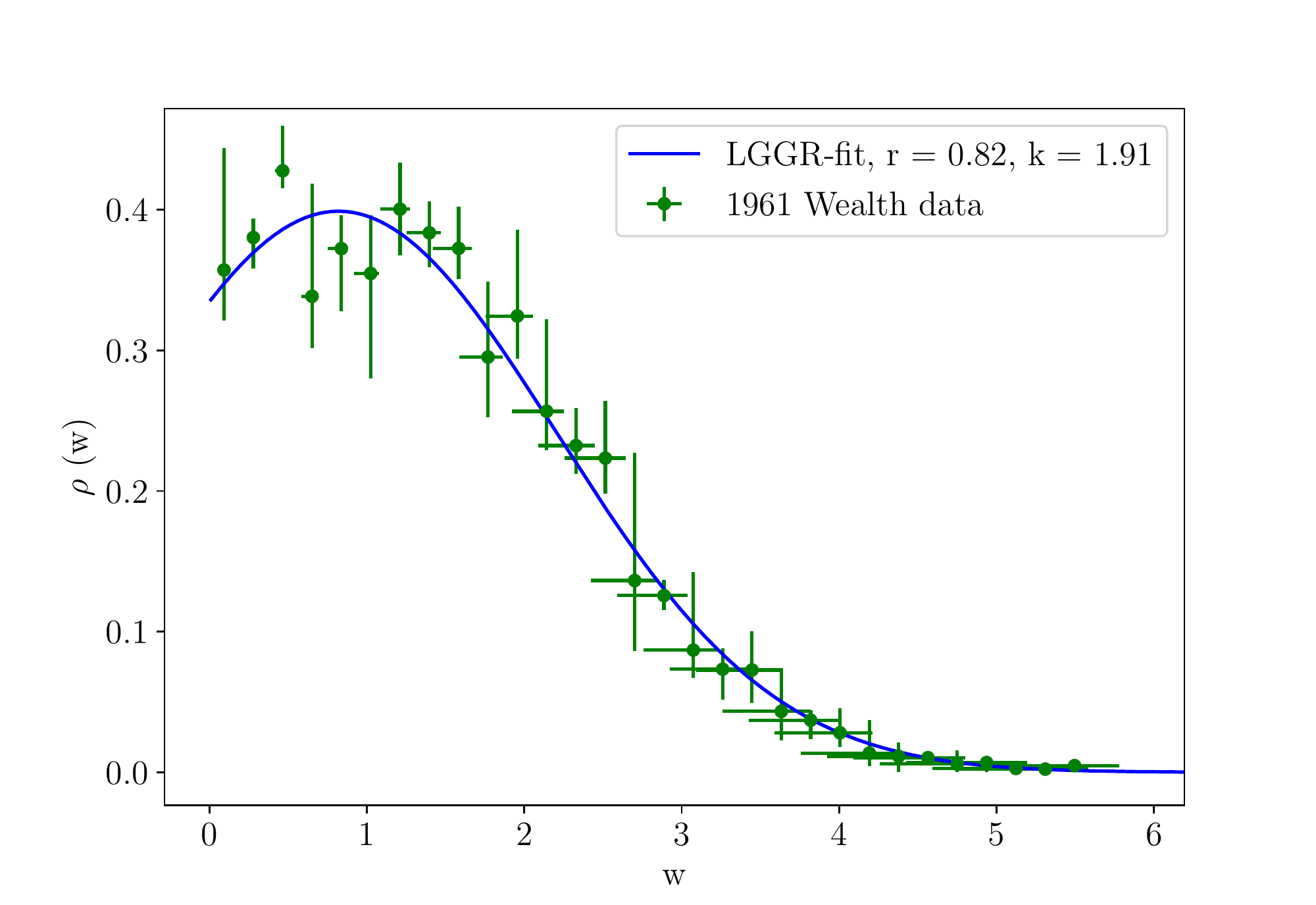}
\end{center}
\caption{Wealth distribution for 1961. Error bars are obtained by combining the results of the different weight parameter sets shown in Table \ref{tbl:Params61} . The theoretical distribution (\ref{pdf-comm}) fitted to the averaged experimental distribution ($k=1.91$ and $r=0.82$) is shown by the continuous line.}
\label{fig:Wealth_dist_1961}
\end{figure} 

The obtained probability density functions are presented in such manner on Figure \ref{fig:Wealth_dist_1961}. These figures suggest that the different  parameter sets introduce some uncertainties affecting the location of the data points, but the overall qualitative shape of the distribution is well delimited.

\subsection{Wealth data for 1989:}
\par The records form the Agricultural register 1989 were digitalized and anonymized.There were in total of $921$ individual households. Before proceeding, one has to note that the composition of the wealth will be very different from the year $1961$. First, this is because the agricultural lands were all fully collectivized  and for a household was allowed a home garden of maximum $40$ acres.  This area also included the surface of the buildings. Secondly new categories appear in the records, such as the owned cars. The methodology for estimating the total wealth is the same as for $1961$.
We assume again that the wealth of a household is the weighted sum of the following  relevant "categories":
\begin{itemize}
 \item $A_L$ - total land area in acres ($a$, $1a = 100m^2$).  Weight parameter: $P_{A_L}$.
 \item $A_{Bh}$ -area of the house owned in $m^2$. Weight parameter: $P_{A_{Bh}}$. 
 \item $A_{Ba}$ - total area of the auxiliary buildings (such as barns, etc.) owned.Weight parameter: $P_{A_{Ba}}$.
 \item $N_{cc}$ - number of carriages. Weight parameter: $P_{N_{cc}}$.
 \item $N_{ca}$ - number of cars owned. Weight parameter: $P_{N_{ca}}$.
 \item $N_C$ - number of the cow. Weight parameter: $P_{N_{C}}$.
 \item $N_{WB}$ - number of domestic water buffaloes. Weight parameter: $P_{N_{WB}}$.
 \item $N_H$ - number of horses. Weight parameter: $P_{N_H}$. 
 \item $N_P$ - the number of pigs. Weight parameter: $P_{N_P}$.
 \item $N_S$ - the number of sheep. Weight parameter: $P_{N_S}$.
\end{itemize}

The used different sets of weigh  parameters are presented in Table \ref{tbl:Params89}. The shares in the total wealth of the different categories, as it was discussed  for $1961$, are presented in Table \ref{tbl:Shares89}.

\begin{figure}[!ht] 
\begin{center}
\includegraphics[width=10cm]{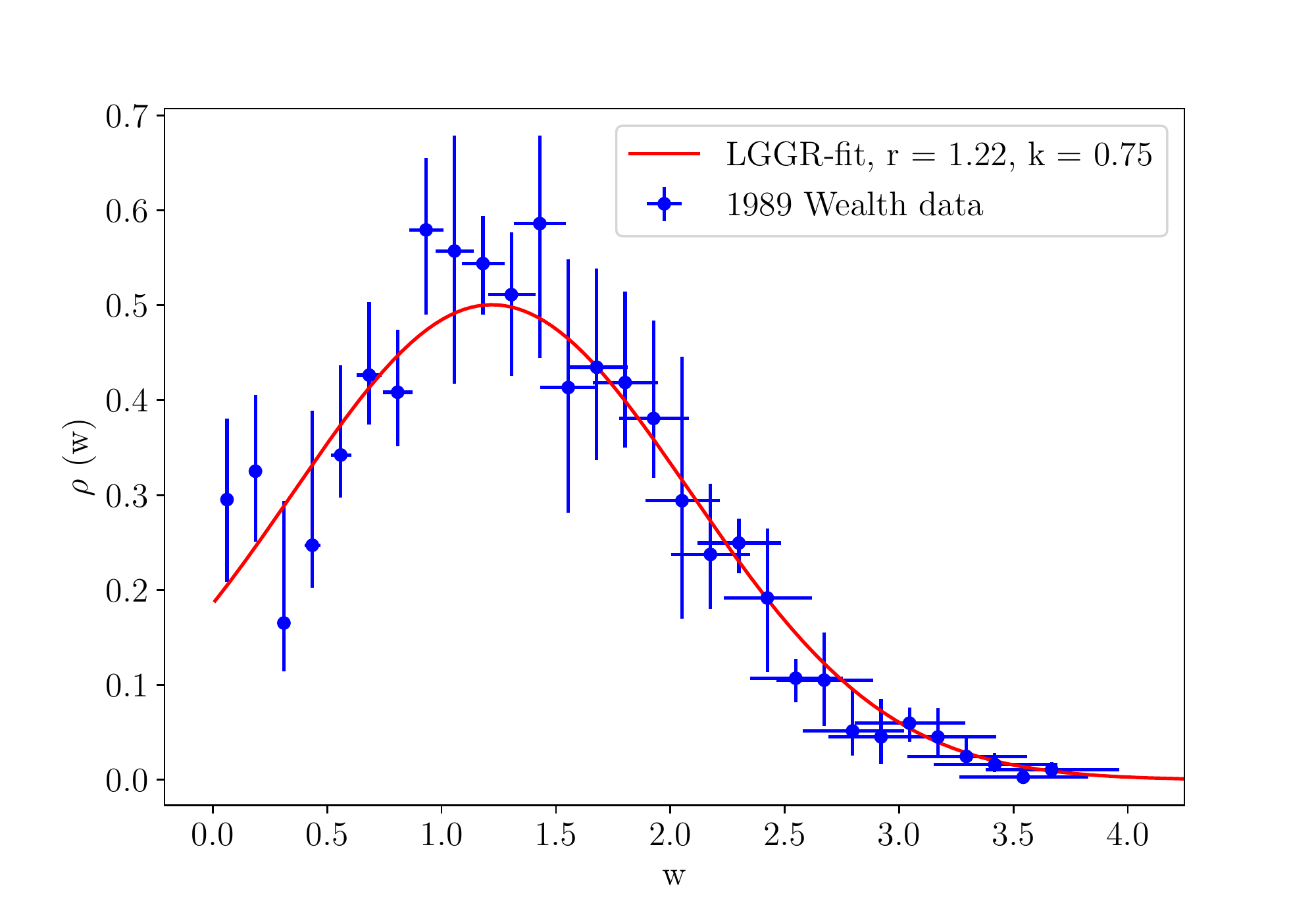}	
\end{center}
\caption{Wealth distribution for 1989. Error bars are obtained by combining the results of the different weight parameter sets shown in Table \ref{tbl:Params89} . The theoretical distribution (\ref{pdf-comm}) fitted to the averaged experimental distribution ($k=0.75$ and $r=1.22$) is shown by the continuous line. }
\label{fig:Wealth_dist_1989}
\end{figure}

The probability densities of wealth distribution $\rho(w)$ was computed in the same manner as for $1961$ and the error bars were estimated again by using the different weight parameter sets.  The result is shown on Figure \ref{fig:Wealth_dist_1989}.

\begin{table}[!ht]
{\small
\begin{center}
\begin{tabular}{|l|l|l|l|l|l|l|l|l|l|l|}
\hline
P. & $P_{A_L}$ & $P_{A_{Bh}}$ & $P_{A_{Ba}}$ & $P_{N_{cc}}$ & $P_{N_{ca}}$ & $P_{N_{C}}$ & $P_{N_{WB}}$ & $P_{N_H}$ & $P_{N_P}$ & $P_{N_S}$ \\ \hline
1  & 0.0116  & 0.0041 & 0.0071 & 0.0747 & 0.5641 & 0.0747 & 0.0747 & 0.1493 & 0.0332 & 0.0066 \\ \hline
2  & 0.0103  & 0.0057  & 0.0063 & 0.0801 & 0.5323 & 0.0801 & 0.0801 & 0.1603 & 0.0372 & 0.0074 \\ \hline
3  & 0.0093 & 0.0069 & 0.0057 & 0.0843 & 0.5083 & 0.0843 & 0.0843 & 0.1686 & 0.0403 & 0.0081 \\ \hline
4  & 0.0085  & 0.0079 & 0.0053 & 0.0876 & 0.4894 & 0.0876 & 0.0876 & 0.1751 & 0.0427 & 0.0085  \\ \hline

5  & 0.0079 & 0.0086 & 0.0049 & 0.0902 & 0.4742 & 0.0902 & 0.0902 & 0.1804 & 0.0446 & 0.0089 \\ \hline
6  & 0.0070 & 0.0098 & 0.0044 & 0.0941 & 0.4512 & 0.0941 & 0.0941 & 0.1883 & 0.0475 & 0.0095 \\ \hline
7  & 0.0066 & 0.0102  & 0.0042 & 0.0957 & 0.4423 & 0.0957 & 0.0957 & 0.1914 & 0.0486 & 0.0097 \\ \hline
8  & 0.0063 & 0.0106  & 0.0040 & 0.0970 & 0.4347 & 0.0970 & 0.0970 & 0.1940 & 0.0496 & 0.0099 \\ \hline
9  & 0.0060 & 0.0109  & 0.0038 & 0.0981 & 0.4280 & 0.0981 & 0.0981 & 0.1963 & 0.0504 & 0.0101  \\ \hline
10 & 0.0058 & 0.0112  & 0.0037 & 0.0991 & 0.4222 & 0.0992 & 0.0992 & 0.1983 & 0.0512 & 0.0102  \\ \hline
\end{tabular}
\end{center}
\caption{Weight parameter sets considered for the data from year 1989.}
\label{tbl:Params89}
}
\end{table}

\begin{table}[!ht]
{\small
\begin{center}
\begin{tabular}{|l|l|l|l|l|l|l|l|l|l|l|}
\hline
P. & $S(A_L)$  & $S(A_{Bh})$ & $S(A_{Ba})$ & $S(N_{cc})$ & $S(N_{ca})$ & $S(N_{C})$ & $S(N_{WB})$ & $S(N_H)$   & $S(N_P)$  & $S(N_S)$   \\ 
  & \% & \%  & \% & \% & \% & \%  & \% & \% & \% & \% \\ \hline
1  & 20.391 & 24.5720  & 41.4460 & 0.52754 & 1.9418 & 1.4000  & 4.2406 & 0.9333 & 4.1692 & 0.3781 \\ \hline
2  & 17.468 & 32.7431 & 35.7360 & 0.54675 & 1.7694 & 1.4510  & 4.3951 & 0.9673 & 4.5140 & 0.4094 \\ \hline
3  & 15.382 & 38.5737 & 31.6616 & 0.56046 & 1.6464 & 1.4874  & 4.5053 & 0.9916 & 4.7600 & 0.4317 \\ \hline
4  & 13.819 & 42.9434 & 28.6081 & 0.57074 & 1.5542 & 1.5147  & 4.5879 & 1.0098 & 4.9443 & 0.4484 \\ \hline
5  & 12.603 & 46.3401 & 26.2345 & 0.57872 & 1.4826 & 1.5359  & 4.6521 & 1.0239 & 5.0876 & 0.4614 \\ \hline
6  & 10.837 & 51.2778 & 22.7841 & 0.59033 & 1.3784  & 1.5667  & 4.7454 & 1.0444 & 5.2960 & 0.4803  \\ \hline
7  & 10.175 & 53.1285 & 21.4908 & 0.59469 & 1.3394 & 1.5782  & 4.7804 & 1.0521 & 5.3740 & 0.4874 \\ \hline
8  & 9.614 & 54.6941 & 20.3968 & 0.59837 & 1.3063 & 1.5880  & 4.8100 & 1.0587 & 5.4401 & 0.4934 \\ \hline
9  & 9.134  & 56.0358 & 19.4593 & 0.60152 & 1.2780 & 1.5964  & 4.8353 & 1.0642 & 5.4967 & 0.4985 \\ \hline
10 & 8.718 & 57.1983 & 18.6469 & 0.60426 & 1.2535 & 1.6036  & 4.8573 & 1.0691 & 5.5456 & 0.5030 \\ \hline
\end{tabular}
\end{center}
\caption{Share in the total wealth for different valuable categories. The rows are the results for the considered parameter sets in year 1989}
\label{tbl:Shares89}
}
\end{table}

\subsection{Wealth data for 2021:}

\begin{figure}[!ht] 
\begin{center}
\includegraphics[width=10cm]{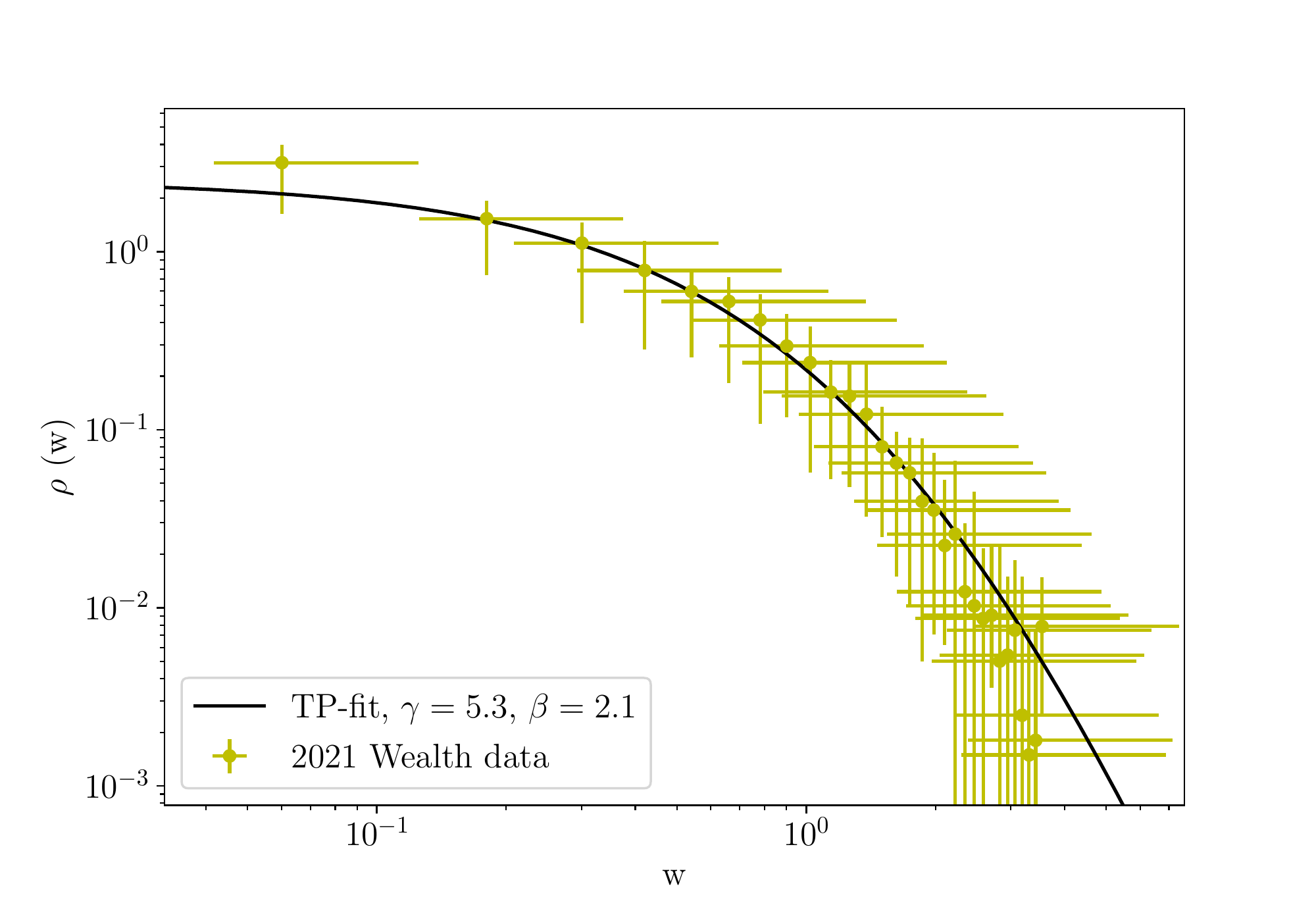}
\end{center}
\caption{Wealth distribution for 2021. Error bars are obtained by combining the results of the different weight parameter sets shown in Table \ref{tbl:Params21} . The theoretical distribution (\ref{pdf-cap}) fitted to the averaged experimental distribution ($\gamma=5.3$ and $\beta=2.1$) 
is shown by the continuous line. 
Please note the logarithmic scales.}
\label{fig:Wealth_dist_2021}
\end{figure}

The wealth distribution for the year 2021 was derived from the anonymized local taxation records. These records contain all the estimated value of the owned real estates, and the size of the owned land inside the villages (possible built area, with running water, electricity and waste management).The land sizes and types owned outside of the village are also present in the database. For $2021$ there are $1595$ taxpaying individuals in these records. The tax an individual pays after their owned valuables are calculated from these records, however we did not have the information regarding payed total tax. Since the values of different land types are not fixed, again realistic weights are needed to combine different assets. The selected "categories" and weights for computing the total wealth of an individual are:
\begin{itemize}
 \item $V_b$ - total estimated value of the owned buildings. It is calculated by taking into account the type of the owned house, the heating apparatus, building materials and etc. by the local authority. Weight parameter: $P_{V_b}$.
 \item $A_{pb}$ - the area of  built-up territories inside the villages in hectares $ha$. Weight parameter: $P_{A_{pb}}$.
 \item $A_{a}$ - the arable field outside the villages in hectares $ha$.Weight parameter: $P_{A_a}$.
 \item $A_{p}$ - the pasture land outside the villages in hectares $ha$. Weight parameter: $P_{A_{p}}$.
 \item $A_g$ - the owned grassland outside the villages in hectares $ha$. Weight parameter: $P_{A_g}$.
 \item $A_f$ - the owned forest outside the villages in hectares $ha$. Weight parameter: $P_{A_f}$.
\end{itemize}
In Table \ref{tbl:Params21} we present the different parameter sets for the weight factors used for 2021. The percentile shares for the different wealth categories calculated according to (\ref{shares}) are presented in Table \ref{tbl:Shares21}.
The methodology of the individual wealth calculation remains the same as in the previous cases. The probability density for wealth distribution with the error bars calculated with the used parameter sets, 
$\rho(w)$,  is given in Figure \ref{fig:Wealth_dist_2021}.

\begin{table}[!ht]
{\small
\begin{center}
\begin{tabular}{|l|l|l|l|l|l|l|}
\hline
P. & $P_{V_b}$   & $P_{A_{pb}}$ & $P_{A_a}$ & $P_{A_p}$ & $P_{A_g}$ & $P_{A_f}$  \\ \hline
1   & 9.5562e-06 & 0.6826   & 0.1502  & 0.0137 & 0.1502  & 0.00341 \\ \hline
2   & 6.6176e-06 & 0.7353   & 0.1176  & 0.0257 & 0.1176  & 0.00368 \\ \hline
3   & 5.1264e-06 & 0.7620   & 0.1011  & 0.0319 & 0.1011  & 0.00381 \\ \hline
4   & 4.2246e-06 & 0.7782   & 0.0912  & 0.0356 & 0.0912  & 0.00389 \\ \hline
5   & 3.6203e-06 & 0.7890   & 0.0845  & 0.0381 & 0.0845  & 0.00394 \\ \hline
6   & 2.8616e-06 & 0.8027   & 0.0761  & 0.0412 & 0.0761  & 0.00401 \\ \hline
7   & 2.6079e-06 & 0.8072   & 0.0733  & 0.0422 & 0.0733  & 0.00403   \\ \hline
8   & 2.4046e-06 & 0.8108   & 0.0710  & 0.0431 & 0.0710  & 0.00405 \\ \hline
9   & 2.2380e-06 & 0.8138   & 0.0692  & 0.0437 & 0.0692  & 0.00406 \\ \hline
10  & 2.0991e-06 & 0.8163   & 0.0676  & 0.0443 & 0.0676  & 0.00408 \\ \hline
\end{tabular}
\end{center}
\caption{Weight parameter sets considered for the data from year 2021.}
\label{tbl:Params21}
}
\end{table}

\begin{table}[!ht]
{\small
\begin{center}
\begin{tabular}{|l|l|l|l|l|l|l|}
\hline
P. & $S(V_b)$  & $S(A_pb)$ & $S(A_a)$  & $S(A_p)$   & $S(A_g)$  & $S(A_f)$    \\ 
   & \% & \% & \% & \% & \% &  \% \\ \hline
1   & 54.3539 & 10.8964 & 24.5545 & 0.1020 & 10.0492 & 0.0440  \\ \hline
2   & 49.0569 & 15.2982 & 25.0717 & 0.2506 & 10.2608 & 0.0618 \\ \hline
3   & 44.9162 & 18.7390 & 25.4759 & 0.3668 & 10.4263 & 0.0757 \\ \hline
4   & 41.5905 & 21.5026 & 25.8006 & 0.4601 & 10.5592 & 0.0869 \\ \hline
5   & 38.8608 & 23.7710 & 26.0672 & 0.5367 & 10.6683 & 0.0961 \\ \hline
6   & 34.6456 & 27.2737 & 26.4787 & 0.6550 & 10.8367 & 0.1102  \\ \hline
7   & 32.9846 & 28.6541 & 26.6409 & 0.7016 & 10.9031 & 0.1158  \\ \hline
8   & 31.5426 & 29.8523 & 26.7817 & 0.7421 & 10.9607 & 0.1207  \\ \hline
9   & 30.2790 & 30.9023 & 26.9051 & 0.7775 & 11.0112 & 0.1249  \\ \hline
10  & 29.1628 & 31.8299 & 27.0140 & 0.8088 & 11.0558 & 0.1287  \\ \hline
\end{tabular}
\end{center}
\caption{Share in the total wealth for different valuable categories. The rows are the results for the considered parameter sets in year 1989}
\label{tbl:Shares21}
}
\end{table}

\section{The LGGR modeling framework}
\begin{figure}[!ht]
\begin{center}
\includegraphics[width=12cm]{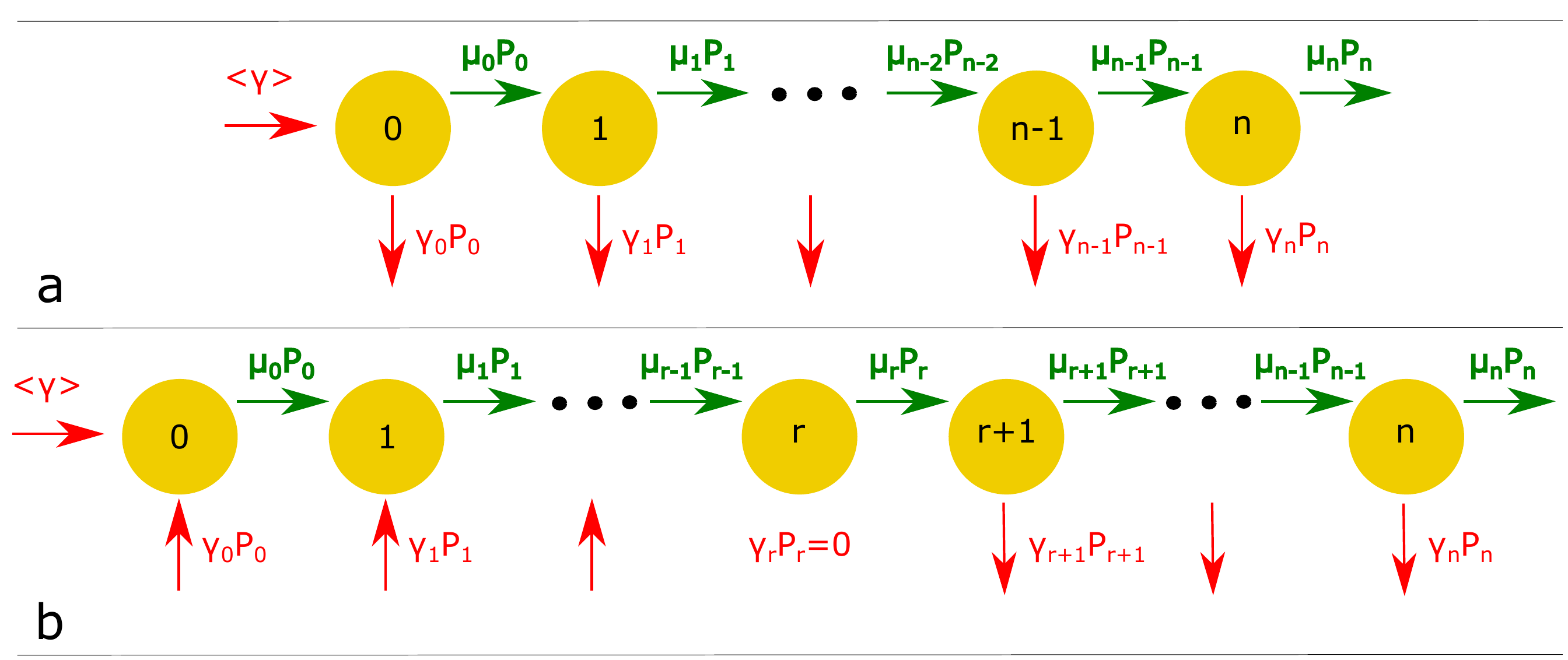}
\end{center}
\caption{Illustration of the growth and reset process: (a) the general mechanism of the process with a positive reset rate, (b) the process considering a reset rate which can be both positive and negative.}
\label{fig:lggr_schem}
\end{figure}

A master equation approach describing uni-directional local growth and global reset processes (LGGR) was recently introduced for modeling various distributions that are frequently encountered in complex systems. The appropriateness of such a simple model to describe income and wealth data in modern societies was also discussed in a couple of recent articles \cite{ZNeda1, ZNeda2}. For a better understanding of the LGGR model let us consider a system composed of identical entities, each of them characterized by the amount of quanta they posses. An immediate example of such a system would be the individuals of a society owning different amount of wealth. Let us denote by $P_n(t)$ the probability that a person has $n$ quanta of wealth at time $t$. The $P_n(t)$ probability has to satisfy normalization: $\sum_{\{n\}} P_n(t)=1$. For some fixed 
$\mu_n$ growth rates and $\gamma_n$ reset rates some possible dynamical scenarios are sketched  in Figure \ref{fig:lggr_schem}. In the  scenario from Figure \ref{fig:lggr_schem}a, the reset rate $\gamma_n$ is positive for all $n$ values. The model allows for a scenario with $\gamma_n$ state-dependent rates where $\gamma_n<0$ if $n<n_c$, and 
$\gamma_n>0$ for $n>n_c$. This would mean as a general rule actors with low wealth $n<n_c$ are coming in the system, and those leaving the system would have $n>n_c$ amount of quanta, in general. The schematic representation of the process in case of this scenario is displayed on Figure \ref{fig:lggr_schem}b. 
In \cite{ZNeda1, ZNeda2} it has been shown, that this second scenario is extremely appropriate to model socio-economic inequalities. Independently, whether we are in the case (a) or (b), the analytical investigation of the LGGR model follows the same route.  

In case when only local unidirectional transitions ($n \rightarrow n+1$) are considered for growth, the dynamical evolution equation will have the following form:

\begin{equation}
\frac{dP_n(t)}{dt}=\mu_{n-1}P_{n-1}(t)-\mu_n P_n(t)-\gamma_nP_n(t) + \delta_{n,0}\langle \gamma \rangle(t) . 
\label{master_dis}
\end{equation} 

The term containing the $\langle \gamma \rangle$ quantity, guarantees the normalization of $P_n(t)$ by feeding the system at the state $n=0$ if needed:
\begin{equation}
\langle \gamma \rangle(t) =\sum_{j} \gamma_j P_j(t).
\end{equation} 
In our previous studies we have demonstrated the generalization of the dynamical Eq. \ref{master_dis} to continuous states by converting it into a partial differential equation in the limit when $dt\rightarrow 0$ \cite{Biro-Neda}. Generalizing the 
growth and reset rates to continuous states ($\mu_n \rightarrow \mu(x)$, $\gamma_n \rightarrow \gamma(x)$), the evolution equation corresponding to Eq. \ref{master_dis} has the following form:
 \begin{equation}
 \frac{\partial \rho(x,t)}{\partial t}=-\frac{\partial}{\partial x} \left[ \mu(x) \rho(x,t) \right] - \gamma(x) \rho(x,t) +\langle \gamma(x) \rangle (t) \delta(x).
 \label{master_gen}
 \end{equation}
 Here $\rho(x,t)$ is the probability density ($\int_{\{x\}} \rho(x,t) dx=1$) for an individual possessing $x$ amount of wealth at time moment $t$.  The feeding term at $0$ is similar to the discrete limit, and it is described with the $\delta(x)$ Dirac functional. The $\langle \gamma \rangle$ feeding at $0$ should be:
 \begin{equation}
 \langle \gamma(x) \rangle (t) = \int_{\{x\}} \gamma(x) \rho(x,t) dx
 \label{normal}
 \end{equation} 
 
 \newcommand{\eadx}[1]{{\rm e}^{#1} }
  In \cite{Biro-Neda,BNT} it was proven, that the above dynamical evolution equation converges to a steady-state with a $\rho_s(x)$ stationary probability density: 
 \begin{equation}
\rho_s(x) \: = \: \frac{\mu(0) \rho_s(0)}{\mu(x)} \, \eadx{-\int_0^x\limits \frac{\gamma(u)}{\mu(u)}du},
\label{stat-distr}
 \end{equation}
By correctly choosing the $\mu(x)$ growth-  and $\gamma(x)$ reset rates,  the LGGR model will  lead to $\rho_s(x)$ distributions that are frequently  encountered in complex systems \cite{Biro-Neda}. 
We will apply thus the LGGR modeling framework to describe the wealth distributions obtained in the mentioned three different economic period of the studied geographical region.  
 
\section{Application of the LGGR model}

Using particularized growth and reset rates we apply now the LGGR model for the three distinct economic situations in order to explain the obtained wealth distributions. First we identify the proper kernels for the reset and growth rates and then calculate the resulting stationary probability density functions. We then adjust the model parameters to obtain a qualitatively acceptable overlap between the experimental and model data.  In a later section we will discuss some aspects related to the chosen growth and reset kernels and derive also some accepted economic indicators of wealth inequalities, like the Gini index, Lorentz curve  or the much-discussed Pareto point. 

As a starting point in our endeavor, we note that unlike to our last work related to wealth inequalities, \cite{ZNeda2}, in our present data there is no information on negative wealth , i.e. debts. As a consequence of this, the application of the LGGR model for describing our data should be 
possible using much simpler kernels for the growth and reset rates. 

\subsection{Wealth distribution in Communism - constant growth and linear reset rates}

For modeling the data in  years 1961 and 1989 we considered the following rates:
\begin{itemize}
 \item a constant growth rate, resulting  in a slow growth independently of the existing wealth amount, in agreement with the principles of communism, 
 $\mu(x) = k$.
 \item a linear reset rate in the form $\gamma(x)=x-r$.  Assuming that only positive wealth exists such a rate will have a negative value in the interval $[0,r)$, and becomes positive on $[r, \infty)$. As discussed in the previous section, we are now in the case illustrated in Figure \ref{fig:lggr_schem}b. A positive reset value means in average exiting from the system at wealth value $x$, while negative reset means in average an entering into the system with wealth $x$. In this manner, the value of the $r$ value marks the $n=n_c$ boundary.  Wealth above this favor a reset, while wealth under it is considered as starting assets for a new individual incoming in the statistics.   
\end{itemize}

Using the above growth and reset rates, equation (\ref{stat-distr}) leads to the stationary probability density:  
\begin{equation}
 \rho_{s1}(x) =\sqrt{\frac{2}{k \pi}}\, e^{\frac{-x(x-2r)}{2k}} \, \left [ erf(\frac{r}{\sqrt{2k}})+1 \right ]^{-1}  
 \label{pdf-comm}
\end{equation}
This is a normalized normal distribution restricted on the $x\ge 0$ interval with its peak shifted into $r$. We denoted here by $erf()$ the 
well-known error-function. By properly selecting the $r$ and $k$ parameters, this approach leads to a good fit for the wealth distributions both for the 1961 and 1989 data. 

For the data and wealth distributions in year \textbf{1961} the best fit can be achieved with $k=1.91$ and $r=0.82$. 
The fit along with the experimental results for the wealth distribution are presented on Figure \ref{fig:Wealth_dist_1961}.
Using the median points for the data, the fit with the above parameters results in a coefficient of determination $R^2 = 0.98$. The analytically calculated average wealth for this normal-like distribution is $\langle x \rangle_{theo} = 1.459 \,a.u.$, which can be compared with the average calculated from the experimental data, for each different parametrization $\langle x \rangle_{exp} \in [1.36\, a.u.;1.49\, a.u.]$. (Here $a.u.$ stands for the arbitrary units in wealth resulting from our estimation method)

The same normal-like distribution offers a good fit also for the wealth distribution derived from the \textbf{1989}  data. The best fit
parameters are however: $k=0.75$ and $r=1.22$. This fit is illustrated on Figure \ref{fig:Wealth_dist_1989}. Using again the 
median points of the experimentally estimated data the above fit leads to a coefficient of determination $R^2 = 0.93$. 
The average wealth from the fitted distribution is $\langle x \rangle_{theo} = 1.359 \,a.u.$, which is in good agreement with the values calculated from our parametrizations of the weight coefficients: $\langle x \rangle_{exp} \in [1.199\, a.u.;1.391\, a.u.]$

From  Figures \ref{fig:Wealth_dist_1961}, \ref{fig:Wealth_dist_1989} and from the fit statistics we learn that the chosen 
LGGR method with the proper $k$ and $r$ parameters describes  well the wealth distribution in the communism periods, more specifically the ones 
derived for 1961 and 1989. We note that the average wealth in this case can be given analytically:
\begin{equation}
 \langle x \rangle = r+\sqrt{\frac{2k}{\pi}} \frac{e^{-\frac{r^2}{2k}}}{[1+erf(\frac{r}{\sqrt{2k}})]}
\end{equation}

For small $k$ values the mean shifts towards the value of $r$, which is consistent with our data and fit for $1961$ and $1989$.

\subsection{Wealth distribution  for the free market - preferential growth with constant reset rate}

As it was shown for the capitalistic free market economy the growth in wealth should be preferential \cite{Biro-Neda}. This leads to the kernel used in our recent study on wealth distribution in modern societies \cite{ZNeda2}. A constant reset rate is the simplest approach, assuming that 
growth starts from zero and there is a constant probability of getting out from the considered statistics (either by relocating or by death).  For modeling the wealth distribution observed in year $2021$ we used thus the LGGR model with the $\mu(x)=x+\beta$ growth rate and $\gamma(x)=\gamma$ reset rate. The preferential growth rate is in agreement with the much discussed Matthew principle \cite{Perc}, while the constant reset rate does not differentiate between wealth values, each individual in the system can be resetted at any given time with the same probability. 

The above growth and reset rates results in equation (\ref{stat-distr}) in the Lomax II type (or Tsallis-Pareto) stationary probability density function:
\begin{equation}
 \rho_{s2}(x) = \frac{\gamma}{\beta} (1+\frac{x}{\beta})^{-1-\gamma}
 \label{pdf-cap}
\end{equation}

For the data from $2021$ we found that the best fit can be achieved with $\gamma=5.3$ and $\beta=2.1$. On Figure \ref{fig:Wealth_dist_2021} we present the experimental probability density function for wealth along with the Tsalis-Pareto distribution obtained with the previously mentioned parameters. The used regression has a coefficient of determination $R^2 = 0.91$. The average wealth computed from the Tsalis-Pareto distribution is 
$\langle x \rangle_{theo} = 0.488\,a.u.$ in comparison with the average wealth obtained from the data: $\langle x \rangle_{exp} \in [0.279 \,a.u.;0.682\,a.u.]$.
 
\section{Socio-economic inequality measures}

We turn now our attention on estimating also the well-known inequality measures used by social sciences. 

We will first construct both the experimental and theoretical Lorenz  curves \cite{Lorenz} calculated from the experimental and fitted probability density functions, respectively. The Lorenz curve indicates the relation between the cumulative share of wealth owned by the households with wealth above $x$, $F(x)$, and the cumulative share of households with wealth greater than $x$, $C(x)$. 
\begin{equation}
 C(x) = \int_x^{\infty} \rho(w) dw
\end{equation}
\begin{equation}
 F(x) = \frac{1}{\langle x \rangle} \int_x^{\infty} w\, \rho(w) dw \\
 \end{equation}
 with:
 \begin{equation}
 \langle x \rangle = \int_0^{\infty} w\, \rho(w)\, dw 
 \end{equation}

One obtains therefore the Lorenz curve by plotting $F(x)$ as a function of $C(x)$. For a totally uniform wealth distribution, i.e. no socio-economic inequality, the Lorenz curve would be the first diagonal ("equity line") in the $F-C$ square. Deviation from this line characterizes the social inequalities. The area between the Lorenz curve and the equity line ($\Gamma$) is related to the well-known Gini index \cite{Gini}, $G$, by $G=2\, \Gamma$. We recall here that if one has a discrete set of wealth values, $x_i$ in a society (the case of our experimental data) composed by $N$ individuals, the Gini index is a number between $0$ and $1$, defined as
\begin{equation}
G=\frac{1}{2 \langle x \rangle}\sum_i^N \sum_j^N \left | x_i-x_j \right| 
\label{gini-discrete}
\end{equation}
with $\langle x \rangle$ the average wealth:
\begin{equation}
\langle x \rangle = \frac{1}{N} \sum_i^N x_i
\end{equation}  
A $0$ Gini index means no social inequality (all wealths are equal) why a Gini index $1$ means that all wealth is 
owned by one individual, i.e. the most extreme inequality.  For a more pronounced social inequality the Gini index is higher. 

Alternatively, if one has a continuous probability density function for the wealth distribution, the Gini index is computable as:
\begin{eqnarray}
G= \frac{1}{\langle x \rangle} \int_0^{\infty} dx \, \int_{x}^{\infty} \, dy \, (y-x) \, \rho(x) \, \rho(y) \label{gini-cont}\\
\langle x \rangle = \int_0^{\infty} x \, \rho(x) \, dx \nonumber
\end{eqnarray}

\begin{figure}[!ht] 
\begin{center}
\includegraphics[width=10cm]{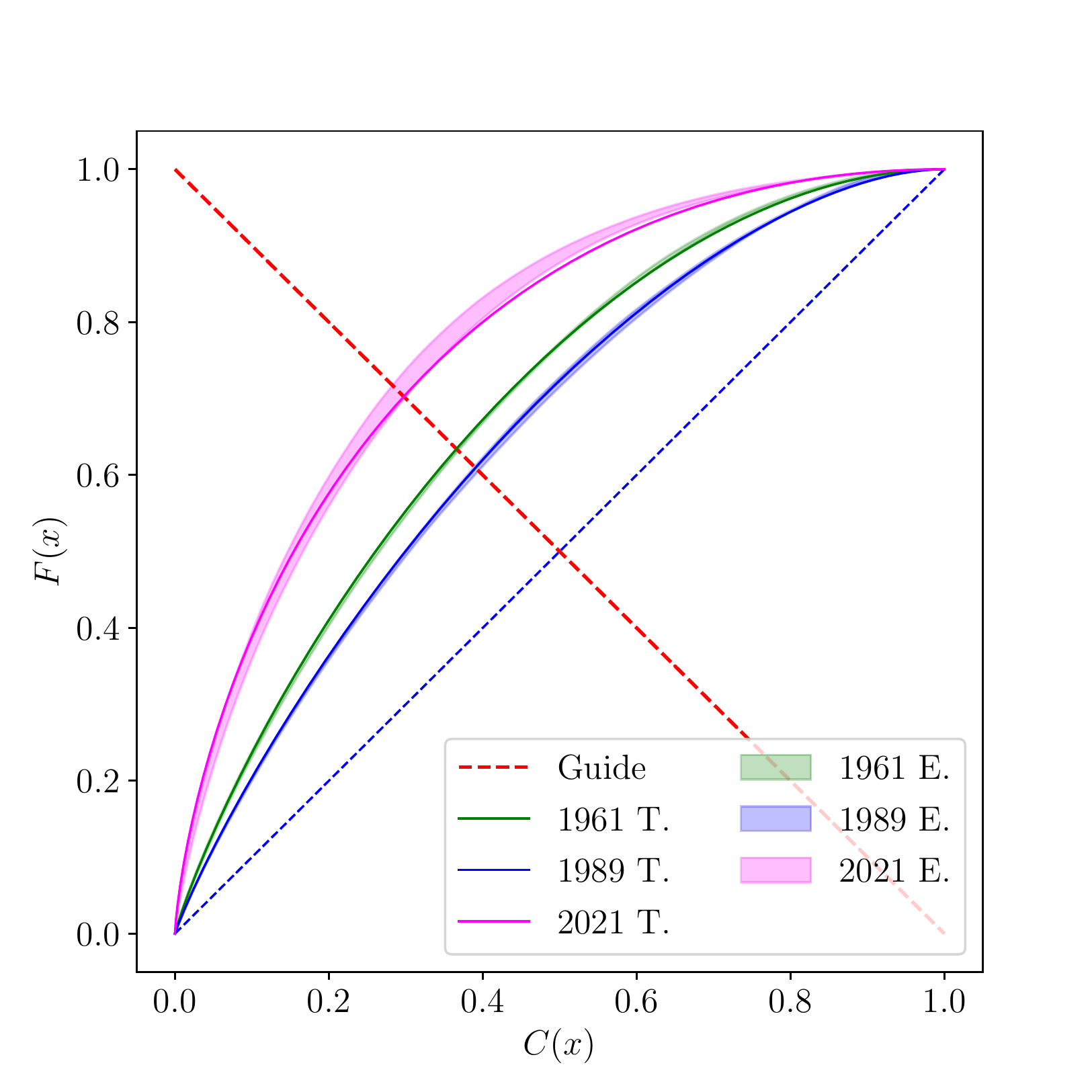}
\end{center}
\caption{Experimental and theoretical Lorenz curves for the years 1961,1989 and 2021.  The shaded region indicates the region spanned by the experimental results for different weight parameter sets. The theoretical curves were calculated from the fitted distributions, and are plotted with the continuous bold line.}
\label{fig:L_curves_all}
\end{figure}

Following the above definitions one can construct both the Lorenz curve and Gini index from the experimental and model results. The Lorentz curves for the studied years are plotted on Figure \ref{fig:L_curves_all}. The regions spanned by the 
experimentally observed Lorenz curves for different weight parameters are indicated with a lighter shedded region. The theoretical Lorenz curve  computed with the fitted probability density is indicated with a bold continuous line.  As expected our theoretical model describes in an acceptable manner also the Lorenz curves. Some deviations are however observed for year $2021$ in the limit of high wealths.

The Gini index can be estimated from the experimental data for all weight parameter sets using the definition (\ref{gini-discrete}). One can compute also a theoretical Gini index using equation (\ref{gini-cont}) with the probability density function given by the LGGR model and best fit parameters. Results in such sense are summarized in the columns corresponding to $G$ in Table \ref{tbl:Inequality}. There is a good agreement between the experimentally calculated and theoretically
computed Gini index, confirming once again the applicability of the theoretically derived probability density functions. 

Another possibility to quantify social inequalities is by defining the $P$ Pareto point. The original Pareto law states that in any
society in general $20\%$ of individuals own $80\%$ of the total wealth. Starting from this hypothesis for any specific wealth distribution data in a society one can define a $P$ Pareto point from the Lorenz curve, assuming that the $P$ point is that $P=C$ value for which $F=1-P$. Naturally, for a society that confirms the 80-20 Pareto law $P=0.2$. The Pareto point should be a number between $0$ and $0.5$, smaller values meaning higher social inequalities.
Our experimental and theoretically constructed Lorenz curves allow the identification of the $P$ Pareto point. From our data we get the Pareto points specified in the columns corresponding to $P$ in Table \ref{tbl:Inequality}. One will note again that similarly with the Gini index the theoretical and experimental values are in good agreement.

\begin{table}[!ht]
{\small
\begin{center}
\begin{tabular}{l|ll|ll|}
\hline
\multicolumn{1}{|l|}{Year} & \multicolumn{2}{c|}{G}                                          & \multicolumn{2}{c|}{P}                                          \\ \hline
                           & \multicolumn{1}{c|}{Exp.}          & \multicolumn{1}{c|}{Theo.} & \multicolumn{1}{c|}{Exp.}          & \multicolumn{1}{c|}{Theo.} \\ \hline
\multicolumn{1}{|l|}{1961} & \multicolumn{1}{l|}{[0.377;0.379]} & 0.378                      & \multicolumn{1}{l|}{[0.366;0.368]} & 0.367                      \\ \hline
\multicolumn{1}{|l|}{1989} & \multicolumn{1}{l|}{[0.304;0.315]} & 0.312                      & \multicolumn{1}{l|}{[0.390;0.395]} & 0.391                      \\ \hline
\multicolumn{1}{|l|}{2021} & \multicolumn{1}{l|}{[0.543;0.579]} & 0.552                      & \multicolumn{1}{l|}{[0.282;0.299]} & 0.298                      \\ \hline
\end{tabular}
\end{center}
\caption{Inequality measures for the studied years: the Gini ($G$) and Pareto-point ($P$). Value limits obtained from the data with different weight parameters and the value obtained from the theoretically fitted probability density function.}
\label{tbl:Inequality}
}
\end{table}

\section{Discussion and Conclusions}
Before discussing in detail the obtained results and derive the usual socio-economic inequality measures,  let us recall here that the chosen weight parameters were crucial in giving a realistic estimate over the total wealth of an individual or household. Looking back to the chosen values of the weight parameters summarized in Tables \ref{tbl:Shares 61}, \ref{tbl:Shares89} and \ref{tbl:Shares21} one will observe that the chosen sets indeed affects the share of wealth categories. 
This uncertainty  will have a direct effect in the wealth averages too, however the error bars from Figures \ref{fig:Wealth_dist_1961} \ref{fig:Wealth_dist_1989} and \ref{fig:Wealth_dist_2021} suggests that the overall shape 
of the probability density functions  are not altered in a qualitative manner. The reason for this is relatively simple. As previously said the commune is primarily agricultural in its economy, meaning that for the year 1961 all the different wealth categories may be considered as proportional to the size of the owned land. Indeed, the number of livestock, the size of the barn and house should all be proportional to it, as the land is needed to support the number of livestock, and dictates the size of a barn owned by a household. In the 1961 social situation the distribution of wealth is already quite egalitarian due
to the land reforms from 1921 and 1945. As a consequence of these, in 1961 the larges owned estate is only $11.76$ hectares of land.

After 1961 all lands were forcefully collectivized, leaving households with a garden, the owned smaller buildings and the house. The maximal size of the garden was limited to $40$ acres, although some larger ones exist in the records (these gardens were located at places where they could not be meaningfully used.) Anecdotal evidence from the commune says that after the collectivization most of the earned yearly income was invested in rebuilding, extending and to building new houses. This is well supported by the data also. While in 1961 the total area of houses owned by households in the commune was $44273m^2$, in 1989 this number became $65396m^2$ which means a $47.7\%$ growth.

As it is observable in Figure \ref{fig:Wealth_dist_2021}, in 2021 one will observe a larger scattering of the probability densities, caused by the modified and uncorrelated wealth components used in the estimation of the total wealth. This is a results of less categories in the taxation database and also reflects that the agriculture based society diversified to other sectors such as tourism. As a result of this second effect the proportionality between the size of the land owned and the other categories may not be true anymore. Clearly, the shape of the distribution is shifted towards the characteristic power-law like tail by the current year.

We found once again that the LGGR model provides a useful modeling base for many complex systems. For the considered  problem in particular, it allowed a realistic modeling of wealth distribution at each historically significant economic period. 
For the communism years (1961 and 1989) the growth rate was chosen as a fixed $k$ constant. A state-dependent growth rate would have raised ideological problems for the communist regime. The constant growth rate leads to a scenario where people may only produce enough for themselves and only trade for the essentials, instead of investing for a greater profit.  The reset rate was considered to be linear, with an $r$ offset. This can be interpreted in the context of a wealth control mechanism, imposing a desired wealth amount, above which a resetting is favored.  Households enter in their wealth evolution dynamics bellow this $r$ value and leave the system over $r$ in average, in agreement with the ideology supported by a communistic regime. As the communist economy evolves our regression results suggests that the growth slows from $k = 1.91$ in 1961 to $k = 0.75$ in 1989, while the reset threshold grows from $r = 0.82$ to $r = 1.22$. The results also suggest that for the smaller $k$ growth rate and the larger $r$ reset threshold the resulting wealth distributions tend to peak at around $r$, with an average wealth also around $r$, as seen in the results for 1989 on Figure \ref{fig:Wealth_dist_1989}.

With  the fall of communism in 1989 the dynamics governing wealth inequalities has changed.  The rules of the open market economy allowed the "rich gets richer"  effect which is modeled here by the linear growth rate. Such a linear growth rate was used in our previous approaches to describe the distribution of income and wealth \cite{ZNeda1,ZNeda2}. The constant reset rate implemented by us means that everybody can leave the system with the same likelihood. The linear growth rate and the constant reset rate in the LGGR model leads to the Tsalis-Pareto stationary distribution. The power-law tail of this distribution is consistent with the accepted experimental results in the current literature. It is interesting to note that the tail of the distribution observed for the studied commune in 2021 is $b = -6.3$ which is much steeper than the tails observed on larger population scales. One can compare this exponent for the one observed in the wealth distributions on country level, where for USA and Russia one gets $b = -2.4$ and France with $b = -2.68$ \cite{ZNeda2}. This steeper decay indicates that the smaller community studied here is much more homogeneous in wealth, and therefore also the socio-economic inequality indicators should be smaller. 

Concerning the inequality measures used in social sciences our results are consistent with the ones expected for the investigated economic periods. The experimentally and theoretically constructed Lorenz curves seems to be in good agreement, confirming from another perspective the validity of the theoretically proposed probability density functions. The experimentally and theoretically calculated $G$ Gini index are also in good agreement, the theoretical values being in the intervals spanned by the experimental values for different parameter sets.  The same observations hold for the value of the $P$ Pareto point. The $G$ and $P$ values in Table \ref{tbl:Inequality}  shows that social inequalities were low in the communist regime. The $G$ value is around $\sim 0.38$ in 1961, decreasing to as low as $\sim 0.31$ in 1989. After 31 years of a market economy the value of $G$ increased in this region to  a value around $\sim 0.55$. The same tendency in the dynamics of social inequalities is observable in the variation of the $P$ Pareto point. The value of $P$ is 
$\sim 0.37$ in 1961 increasing to a value $\sim 0.39$ in 1989. The value of $P$ nowadays in this region is $\sim 0.3$, indicating the deepening social inequalities. 

In summary we can state that the experimental and theoretical investigation of wealth distribution in a well-delimited social environment (traditional commune in Transylvania) for economically different situations proved expected variations in 
social inequality measures. Along with obtained very precious inequality data, the study proves once again the modeling power of the LGGR model in various interdisciplinary problems related to complex systems.

\section*{Conflict of Interest Statement}
The authors declare that the research was conducted in the absence of any commercial or financial relationships that could be considered as a potential conflict of interest.

\section*{Author Contributions}
Conceptualization and research design by ZN; theoretical model by TB and ZN; data mining by IG;  data analysis by IG and SzK; Figures by IG and SzK; interpretation by ZN and TB; first draft of the manuscript by IG, SzK and ZN; all author contributed to the final form of the manuscript.  
 
\section*{Funding}
Work supported by the UEFISCDI  PN-III-P4-ID-PCE-2020-0647 research grant. T.S. Bir\'o
has been also supported by the Hungarian National Bureau for Research, Innovation and Development under project Nr. K123815. The work of Sz. Kelemen is also supported by the Collegium Talentum Program of Hungary.

\section*{Acknowledgments}
We acknowledge the mayor office of the Sancraiu commune for the help and openness in providing us the Agricultural records from the archives, and the help they gave for the extraction of the anonymized taxation data. We also acknowledge senior Istv\'an Gere for the enlightening suggestions on the data sources.

\section*{Data Availability Statement}
The datasets curated and analysed for this study can be found in the Figshare repository: "Wealth distribution in villages - Datasets"  \url{https://www.doi.org/10.6084/m9.figshare.17013467}.

%Bibliography
\bibliographystyle{unsrt}  
\bibliography{references}

\begin{thebibliography}{10}

\bibitem{scalingincities}
L.M.A. Bettencourt.
\newblock The origins of scaling in cities.
\newblock {\em Science}, 340(6139):1438--1441, 2013.

\bibitem{scalingincities3}
E.~Heinrich~Mora, C.~Heine, J.J. Jackson, G.B. West, V.C. Yang, and C.P.
  Kempes.
\newblock Scaling of urban income inequality in the usa.
\newblock {\em Journal of the Royal Society Interface}, 18(181):20210223, 2021.

\bibitem{scalingincities4}
E.~Arcaute, E.~Hatna, P.~Ferguson, H.~Youn, A.~Johansson, and M.~Batty.
\newblock Constructing cities, deconstructing scaling laws.
\newblock {\em Journal of the Royal Society Interface}, 12(102):20140745, 2015.

\bibitem{scalingincities5}
L.M.A. Bettencourt, J.~Lobo, D.~Helbing, C.~K{\"u}hnert, and G.B. West.
\newblock Growth, innovation, scaling, and the pace of life in cities.
\newblock {\em Proceedings of the National Academy of Sciences},
  104(17):7301--7306, 2007.

\bibitem{scalingincities6}
A.F.J. Van~Raan, G.~Van Der~Meulen, and W.~Goedhart.
\newblock Urban scaling of cities in the netherlands.
\newblock {\em PLoS One}, 11(1):e0146775, 2016.

\bibitem{scalingincities7}
K.~Behrens and F.~Robert-Nicoud.
\newblock Survival of the fittest in cities: Urbanisation and inequality.
\newblock {\em The Economic Journal}, 124(581):1371--1400, 2014.

\bibitem{scalingincities8}
J.~Depersin and M.~Barthelemy.
\newblock From global scaling to the dynamics of individual cities.
\newblock {\em Proceedings of the National Academy of Sciences},
  115(10):2317--2322, 2018.

\bibitem{chakraborti}
A.~Chakraborti, A.~Chatterjee, B.~Chakrabarti, and S.R. Chakravarty.
\newblock {\em Econophysics of Income and Wealth Distributions}.
\newblock Cambridge Univ. Press, Cambridge, 2013.

\bibitem{pareto}
V.~Pareto.
\newblock Cours d’economie politique, vol. 2.
\newblock {\em Pichou, Paris}, 1897.

\bibitem{yakovenko}
A.~Dr{\u{a}}gulescu and V.M. Yakovenko.
\newblock Evidence for the exponential distribution of income in the usa.
\newblock {\em The European Physical Journal B-Condensed Matter and Complex
  Systems}, 20(4):585--589, 2001.

\bibitem{yakovenkko09}
V.M. Yakovenko and J.B. Rosser~Jr.
\newblock Colloquium: Statistical mechanics of money, wealth, and income.
\newblock {\em Reviews of Modern Physics}, 81(4):1703, 2009.

\bibitem{ZNeda2}
I.~Gere, Sz. Kelemen, G.~T{\'o}th, T.S. Bir{\'o}, and Z.~N{\'e}da.
\newblock Wealth distribution in modern societies: collected data and a master
  equation approach.
\newblock {\em Physica A: Statistical Mechanics and its Applications}, page
  126194, 2021.

\bibitem{Cui}
L.~Cui and C.~Lin.
\newblock A simple and efficient kinetic model for wealth distribution with
  saving propensity effect: Based on lattice gas automaton.
\newblock {\em Physica A: Statistical Mechanics and its Applications},
  561:125283, 2021.

\bibitem{Cardoso}
B.H.F. Cardoso, S.~Gon{\c{c}}alves, and J.R. Iglesias.
\newblock Wealth distribution models with regulations: Dynamics and equilibria.
\newblock {\em Physica A: Statistical Mechanics and its Applications},
  551:124201, 2020.

\bibitem{Jongsoon}
P.~Jongsoon and P.~Youngho.
\newblock Wealth distribution for the spin agent model of the stock market.
\newblock {\em New Physics: Sae Mulli}, 2020.

\bibitem{Lim}
G.~Lim and S.~Min.
\newblock Analysis of solidarity effect for entropy, pareto, and gini indices
  on two-class society using kinetic wealth exchange model.
\newblock {\em Entropy}, 22(4):386, 2020.

\bibitem{levy}
M.~Levy and H.~Levy.
\newblock Investment talent and the pareto wealth distribution: Theoretical and
  experimental analysis.
\newblock {\em Review of Economics and Statistics}, 85(3):709--725, 2003.

\bibitem{jones}
C.I. Jones.
\newblock Pareto and {P}iketty: The macroeconomics of top income and wealth
  inequality.
\newblock {\em Journal of Economic Perspectives}, 29(1):29--46, 2015.

\bibitem{sorger}
G.~Sorger.
\newblock Income and wealth distribution in a simple model of growth.
\newblock {\em Economic Theory}, 16(1):23--42, 2000.

\bibitem{Clementi}
F.~Clementi, M.~Gallegati, and G.~Kaniadakis.
\newblock A generalized statistical model for the size distribution of wealth.
\newblock {\em Journal of Statistical Mechanics: Theory and Experiment},
  2012(12):P12006, 2012.

\bibitem{Bouchaud}
J.P. Bouchaud and M.~M{\'e}zard.
\newblock Wealth condensation in a simple model of economy.
\newblock {\em Physica A: Statistical Mechanics and its Applications},
  282(3-4):536--545, 2000.

\bibitem{Chatterjee}
A.~Chatterjee, B.K. Chakrabarti, and R.B. Stinchcombe.
\newblock Master equation for a kinetic model of a trading market and its
  analytic solution.
\newblock {\em Physical Review E}, 72(2):026126, 2005.

\bibitem{Oliveira}
P.M.C. de~Oliveira.
\newblock Investment/taxation/redistribution model criticality.
\newblock {\em The European Physical Journal B}, 93(10):1--6, 2020.

\bibitem{ZNeda_gambler}
Z.~N{\'e}da, L.~Davidova, Sz. {\'U}jv{\'a}ri, and G.~Istrate.
\newblock Gambler’s ruin problem on erd{\H{o}}s--r{\'e}nyi graphs.
\newblock {\em Physica A: Statistical Mechanics and its Applications},
  468:147--157, 2017.

\bibitem{ZNeda_family}
R.~Coelho, Z.~N{\'e}da, J.J. Ramasco, and M.A. Santos.
\newblock A family-network model for wealth distribution in societies.
\newblock {\em Physica A: Statistical Mechanics and its Applications},
  353:515--528, 2005.

\bibitem{ZNeda1}
Z.~N{\'e}da, I.~Gere, T.S. Bir{\'o}, G.~T{\'o}th, and N.~Derzsy.
\newblock Scaling in income inequalities and its dynamical origin.
\newblock {\em Physica A: Statistical Mechanics and its Applications},
  549:124491, 2020.

\bibitem{Ciesla}
M.~Cie{\'s}la and M.~Snarska.
\newblock A simple mechanism causing wealth concentration.
\newblock {\em Entropy}, 22(10):1148, 2020.

\bibitem{BNT}
T.S. Bir{\'o}, Z.~N{\'e}da, and A.~Telcs.
\newblock Entropic divergence and entropy related to nonlinear master
  equations.
\newblock {\em Entropy}, 21(10):993, 2019.

\bibitem{income_n}
N.~Derzsy, Z.~N{\'e}da, and M.A. Santos.
\newblock Income distribution patterns from a complete social security
  database.
\newblock {\em Physica A: Statistical Mechanics and its Applications},
  391(22):5611--5619, 2012.

\bibitem{yakovenko_w}
A.~Dr{\u{a}}gulescu and V.M. Yakovenko.
\newblock Exponential and power-law probability distributions of wealth and
  income in the united kingdom and the united states.
\newblock {\em Physica A: Statistical Mechanics and its Applications},
  299(1-2):213--221, 2001.

\bibitem{census_older}
E.\'A. Varga.
\newblock {\em Erd\'ely etnikai \'es felekezeti statisztik\'aja. IV. Feh\'er,
  Beszterce-Nasz\'od \'es Kolozs megye. N\'epsz\'aml\'al\'asi adatok 1850-1992
  k\"{o}z\"{o}tt}.
\newblock Pro-Print, Cs\'ikszereda, 2001.

\bibitem{Biro-Neda}
T.S. Bir{\'o} and Z.~N{\'e}da.
\newblock Unidirectional random growth with resetting.
\newblock {\em Physica A: Statistical Mechanics and its Applications},
  499:335--361, 2018.

\bibitem{Perc}
M.~Perc.
\newblock The {M}atthew effect in empirical data.
\newblock {\em Journal of the Royal Society Interface}, 11:20140378, 2014.

\bibitem{Lorenz}
M.O. Lorenz.
\newblock Methods of measuring the concentration of wealth.
\newblock {\em Publications of the American Statistical Association},
  9:209--219, 1905.

\bibitem{Gini}
C.~Gini.
\newblock On the measure of concentration with special reference to income and
  statistics.
\newblock {\em Colorado College Publication, General Series}, 208:73--79, 2001.

\end{thebibliography}

\end{document}